# Superconductivity up to 243 K in yttrium hydrides under high pressure

P. P. Kong[1], V. S. Minkov[1], M. A. Kuzovnikov[1,5], S. P. Besedin[1], A. P. Drozdov[1], S. Mozaffari[2], L. Balicas[2], F.F. Balakirev[3], V. B. Prakapenka[4], E. Greenberg[4], D. A. Knyazev[1] and M. I. Eremets[1*]

[1]*Max-Planck Institut für Chemie, Hahn-Meitner Weg 1, 55128 Mainz, Germany*
[2]*National High Magnetic Field Laboratory, Florida State University, Tallahassee, Florida 32310, USA*
[3]*NHMFL, Los Alamos National Laboratory, MS E536, Los Alamos, New Mexico 87545, USA*
[4]*Center for Advanced Radiation Sources, University of Chicago, 5640 South Ellis Avenue, Chicago, Illinois, 60637, USA*
[5]*Institute of Solid State Physics Russian Academy of Sciences, 2 Academician Ossipyan str., Chernogolovka, Moscow District 142432, Russia*

The discovery of high-temperature conventional superconductivity in H$_3$S with a critical temperature of $T_c$=203 K[1] was followed by the recent record of $T_c$ ~250 K in the face-centered cubic (*fcc*) lanthanum hydride LaH$_{10}$[2,3] compound. It was realized in a new class of hydrogen-dominated compounds having a clathrate-like crystal structure in which hydrogen atoms form a 3D framework and surround a host atom of rare earth elements[4,5]. Yttrium hydrides are predicted to have even higher $T_c$ exceeding room temperature. In this paper, we synthesized and refined the crystal structure of new hydrides: YH$_4$, YH$_6$, and YH$_9$ at pressures up to 237 GPa finding that YH$_4$ crystalizes in the *I4/mmm* lattice, YH$_6$ in *Im-3m* lattice and YH$_9$ in *P6$_3$/mmc* lattice in excellent agreement with the calculations[5,6]. The observed very high-temperature superconductivity is comparable to that found in *fcc*-LaH$_{10}$[2]: the pressure dependence of $T_c$ for YH$_9$ also displays a "dome like shape" with the highest $T_c$ of 243 K at 201 GPa. We also observed a $T_c$ of 227 K at 237 GPa for the YH$_6$ phase. However, the measured $T_c$s are notably lower by ~30 K than predicted[5,6]. Evidence for superconductivity includes the observation of zero electrical resistance, a decrease of $T_c$ under an external magnetic field and an isotope effect. The theoretically predicted *fcc* YH$_{10}$ with the promising highest $T_c$>300 K was not stabilized in our experiments under pressures up to 237 GPa.

**Introduction**

One of the key characteristics of superconductivity is a critical temperature ($T_c$) below this temperature a metal can undergo an electronic transition towards a zero resistance state. Room temperature superconductors (RTSCs) have the potential to revolutionize our world. Perhaps the most straightforward way to reach RTSC can be found within the framework of conventional superconductivity, where the theory is well established. In principle, $T_c$ can be high in a metal displaying suitable parameters: an appropriate high-frequency phonon spectrum and a strong interaction between electrons and phonons at the Fermi level. The formula derived from the Bardeen-Cooper-Schrieffer and Migdal-Eliashberg theories put no apparent limit in $T_c$ [7]. $T_c$, however, depends on a number of factors, such as the details of the phonon and electronic spectra, which are difficult to estimate. Therefore, the theory provides only a general suggestion for the search of particular systems. Such systems should have high characteristic phonon frequencies; for example, it can be realized with light atoms and strong covalent bonding. Based on these criteria, superconductivity was found in $MgB_2$[8] but with a modest $T_c$ of 39 K. The only material which has been always considered as a potential room-temperature superconductor is metallic hydrogen[9]. Although superconductivity in metallic hydrogen has not yet been experimentally observed, this challenge has stimulated the idea to explore other hydrogen containing compounds such as $CH_4$, $SiH_4$, etc., as plausible candidates for the high $T_c$s.[10] These compounds are insulators under ambient conditions, but they can be converted into covalently-bonded metals with the aid of accessible pressures, that are much lower than those required for the metallization of pure hydrogen. While high $T_c$s have not been found in the initially suggested hydrogen-dominant materials[10], the approach itself has proven to be very successful: superconductivity with $T_c$=203 K[1] was found in hydrogen sulfide, despite a modest hydrogen content in this compound. Further theoretical and experimental research in the $H_3S$ related family has not revealed any compound with a higher $T_c$. For instance, hydrogen selenide has a substantially lower $T_c$ of ~105 K[11]. The discovery of new metal superhydrides with the so-called clathrate-like structure of hydrogen atoms raised the calculated $T_c$s close to the room temperature or even above it. In these hydrides, such as in the very first predicted $CaH_6$ with a $T_c$ of 235 K[12], hydrogen atoms create a cage around the host calcium atom. Being connected to each other in the network through weak covalent bonding, hydrogen atoms only weakly interact with the host metal atom through ionic bonding. The host atom supplies a charge to the hydrogen system, stabilizes the crystal structure and promotes metallization at a pressure below 200 GPa. The hydrogen network with short H-H distances in the clathrate-like structures is even closer to metallic atomic hydrogen in comparison to $H_3S$, and can be considered as doped atomic hydrogen. The rapidly growing literature on this topic indicates that various transition metal, lanthanide or actinide elements are prone to form such superhydrides and some of them exhibits superconductivity with much higher calculated critical temperatures[4-6,12-15]. The experimental evidence for the record $T_c$ of ~250 K in the *fcc*-LaH$_{10}$ at 150 GPa[2,3] has confirmed the theoretical predictions and inspired experimentalists to synthesize new members of the clathrate-like family of hydrides with promising high $T_c$s. In the present work, we studied the yttrium-hydrogen system, which is the most attractive candidate for very high $T_c$s among all binary metal-hydrogen systems theoretically studied so far. According to the calculations, superconducting *fcc*-YH$_{10}$ should have extremely high $T_c$, that is as high as 303 K at 400 GPa[5] or 305−326 K at 250 GPa[13]. In addition to YH$_{10}$, there are other phases predicted to display very high $T_c$s and to be stable at lower and more accessible pressures: *hcp*-YH$_9$ with a $T_c$=253−276 K can be stabilized at $P$ = 200 GPa[5], and *bcc*-YH$_6$ with $T_c$=251−264 K that is stable already at 110 GPa[6]. In our experiments, we found the superconducting phases YH$_6$ and YH$_9$ in agreement with the predicted structures[5,6], but with $T_c$s significantly lower than prediction by ~30 K.

## Results and Discussion

Under ambient conditions, the yttrium hydride with highest hydrogen content is *hcp*-YH$_3$. It is a black narrow-bandgap semiconductor with a metallic luster. When compressed in an inert medium, *hcp*-YH$_3$ converts to *fcc*-YH$_3$[16]. This pressure-induced phase transformation is extended under a wide pressure range of 10−25 GPa[17,18]. Further increase of pressure causes likely continuous metallization above 50 GPa[19], as evidenced by the disappearance of the Raman spectrum and a significant drop in resistance.

*Fcc*-YH$_3$ was predicted to be stable under high pressures up to 197 GPa[20]. We found that YH$_3$ and YD$_3$ retain the *fcc* structure of the metallic lattice upon compression without medium up to $P$=180 GPa (Samples 13, 15-18, SM Table 1) and do not exhibit the superconductive transition when subjected to pressures up to 170 GPa upon cooling down to 5 K. For these samples, we observed the appearance of a new *fcc* phase in addition to *fcc*-YH(D)$_3$ under $P$ =130–180 GPa. This new phase atom lattice volume is smaller, than that for *fcc*-YH(D)$_3$ by ~5 Å$^3$/Y, likely indicating reduced hydrogen content. A similar phenomenon was reported earlier upon compression of a substoichiometric LaH$_{3-x}$ in an inert medium[21], where it was interpreted as resulting from a disproportionation reaction into hydrogen-rich stoichiometric trihydride and a hydrogen-poor solid solution. Given that our initial yttrium hydride samples were also substoichiometric, the appearance of the hydrogen-depleted *fcc* phase in dry compression runs could also result from a disproportionation reaction.

When we compressed Y+H$_2$ and Y+D$_2$ mixtures, we observed the formation of YH$_3$ and YD$_3$ already at 17 GPa (e.g. in sample 8), judging from a comparison between the atomic volume measured by XRD and the literature data for YH$_3$[18]. In such experiments, we did not observe the hydrogen-depleted *fcc* phase with a smaller unit cell parameter, which indicates a complete chemical reaction into the saturated stoichiometric yttrium trihydride under excess hydrogen conditions.

The electrical properties and the structures of the yttrium hydrides with a H stoichiometry higher than YH(D)$_3$ were of particular interest in the present study. Such hydrides were synthesized directly from mixtures of Y (samples 3−5), YH$_3$ (samples 1 and 2) or YD$_3$ (samples 6 and 7) in excess of hydrogen (deuterium) under high pressures. The chemical reaction occurs already at room temperature after several weeks, but can be significantly accelerated with a pulsed laser that can heat the mixture up to ~2000 K. In particular, the *I4/mmm* YH$_4$ and *Im-3m* YH$_6$ were synthesized at a pressure of ~160 GPa with the aid of pulsed laser heating up to ~1000 K (samples 4 and 5 in SM Table 1). After several weeks under higher pressures between ~200 and 237 GPa, YH$_4$ and YH$_6$ can be synthesized already at room temperature (samples 1 and 2 in SM Table 1). The *P6$_3$/mmc* YH$_9$ phase can be synthesized starting from *P*~184 GPa but only with the aid of pulsed laser heating (samples 1, 2 and 3, see details in SM Table 1). Higher pressures apparently promote phase transformation: YH$_9$ is synthesized at 237 GPa even upon a subtle sample heating kept below 700 K (no glowing observed) for sample 1 (see the structural details in Fig.2).

Samples prepared from YH$_3$+H$_2$ mixtures show much sharper transitions and a perfect zero resistance state (see the details in Fig.1 (a)). For example, sample 1, which corresponds to the *Im-3m* phase of YH$_6$, was compressed to 237 GPa and kept at room temperature for 3 weeks, showed a sharp transition at 227 K to a zero resistance state (blue curve in Fig.1 (a)).For structural determination see Figs.2 (d) and 2 (f). After keeping sample 2 under 201 GPa at room temperature for ~1 month, a $T_c$ of 210 K in the *Im-3m* YH$_6$ phase was observed. With the aid of pulsed laser heating, the *Im-3m* YH$_6$ phase was transformed into the *P6$_3$/mmc* YH$_9$ phase with a $T_c$ of 243 K (red curve in Fig. 1 (a), see Figs.2 (b) and 2 (g) for the details concerning the identification of the structures). The superconductivity with a $T_c$ of 243 K could be ascribed to YH$_9$, as samples 1 and 2 displayed $T_c$ values around 210−220 K before the pulsed laser heating (or in the absence of the YH$_9$ phase), and after the pulsed laser heating, the *P6$_3$/mmc* YH$_9$ phase was observed with $T_c$ increasing to ~240 K. The main impurity phase in samples 1 and 2 before laser heating was the *I4/mmm* YH$_4$ phase with $c/a \approx 1.9$. This *I4/mmm* YH$_4$ phase was found in many of our samples (1−5), and its XRD pattern is shown in Fig.2 (a). Presently, we are unable to produce reasonably pure *I4/mmm* YH$_4$ to study its superconducting properties. According to calculations[6], the *I4/mmm* YH$_4$ phase is superconducting with a $T_c$=84−95 K, which is considerably lower than those measured in our samples and $T_c$s in the range 251−264 K, predicted for *bcc*-YH$_6$. Thus, superconductivity with a $T_c$ of 227 K and 210 K, as observed in samples 1 and 2 respectively, could be ascribed to the *bcc*-YH$_6$ phase.

The pressure dependence of $T_c$ for the *P6$_3$/mmc* YH$_9$ and the *Im-3m* YH$_6$ phases from different samples is summarized in Figs.1 (b) and 1 (c). It is clearly seen in Fig.1 (b) that the pressure dependence of $T_c$ for YH$_9$ has a "dome like shape" with the highest $T_c$ at ~243 K under 201 GPa, which is similar to the value found for *fcc*-LaH$_{10}$[2]. The range of stability of the YH$_9$ phase differs from the prediction[5] – this phase is stable at lower pressures. We found that the *P6$_3$/mmc* YH$_9$ phase with higher $T_c$ becomes stable from at least 185 GPa. The unexpected abrupt drop in $T_c$(P) in the pressure range of 170−185 GPa, as shown by the open black circles in Fig.1 (b), is probably related to the continuous distortion in the crystal lattice as observed in SH$_3$ at pressures below ∼150 GPa[1]. In Fig.1 (c), for samples 1 and 2 (mixture of *Im-3m* YH$_6$ and *I4/mmm* YH$_4$ phases), the onset $T_c$s were ascribed to the *Im-3m* YH$_6$ phase. For sample 2, we defined the main superconducting transition at 210 K. A small drop in resistance was also observed at 220 K, which is indicated by the smaller open black squares in Fig. 1 (c). However, sample 4 demonstrated a higher $T_c$∼220 K with respect to sample 2. In addition to the *Im-3m* YH$_6$ and the *I4/mmm* YH$_4$ phases, sample 4 also contained an unidentified complex phase (or a mixture of phases). Because the crystal structure and the stoichiometry of the impurities are not determined, it is not clear whether the superconductivity observed in sample 4 is attributable to the YH$_6$ phase. Recently, Troyan *et al*.[22] observed superconductivity in the yttrium hydrides, synthesized through laser heating yttrium and ammonia borane under high pressures. Similarly to sample 4, their samples, revealed a $T_c$ of 224 and 218 K at 165 GPa, and comprised a complex mixture of phases; including the *bcc*-YH$_6$, two new phases with claimed compositions YH$_7$, Y$_2$H$_{15}$ and an unidentified impurity phase. On the basis of *ab-initio* calculations, Troyan *et al*. concluded that these phases should have lower $T_c$s, and assigned $T_c$ ∼220 K to *bcc*-YH$_6$. However, the poor agreement between the experimentally observed XRD patterns and the proposed structural models (YH$_7$ and Y$_2$H$_{15}$ phases) raises concerns about the reliability of their interpretation. Therefore, the superconducting properties of the pure YH$_6$ phase at 160−200 GPa remain open.

Besides the observed drops in resistance to zero value upon cooling, superconductivity was verified by the application of external magnetic fields up to $\mu_0 H = 9$ T. Figure 1 (d) illustrates the dependence of the superconducting transition from sample 5 on an external magnetic field. Upper critical fields as a function of temperature following the criterium of 90% of the resistance in the metallic state are shown in the inset of Fig. 1(d). The solid curve in the inset is an extrapolation to estimate the upper critical fields in the limit of zero-temperatures, after fitting the experimental data to the empirical relation: $H_{c2}(T) = H_{c2}(0)\left(1 - \left(T/T_c\right)^2\right)$. This extrapolation yields $H_{c2}(0)$ =110 T which is about 20 T larger than $H_{c2}(0)$ value in H$_3$S [23]. The zero resistance for the phases YH$_6$ and YH$_9$ (Fig.1 (a)) as well as the characteristic shift of $T_c$ as a function of the magnetic field (Fig.1 (d)) is a clear indication for superconductivity.

To determine the superconducting pairing mechanism, we substituted hydrogen with deuterium to evaluate the effect on $T_c$. We observed a superconductivity with a $T_c$ of 170 K in sample 6 which was synthesized through the pulsed laser heating of $YD_3$ under deuterium atmosphere at temperatures >1000 K and a pressure of 170 GPa. The structural determination still is in progress.

The crystallographic structure of all samples exhibiting superconducting transitions was determined with the aid of powder X-ray diffraction. The Rietveld refinement for the *I4/mmm* $YH_4$, *Im-3m* $YH_6$, and *P6_3/mmc* $YH_9$ crystal structures are shown in Figs. 2 (a), 2 (b), 2 (c), respectively. Figures 2 (d) and 2 (e) are cake representations of the raw X-ray diffraction patterns collected for sample 1 before and after the pulsed laser heating. Figs. 2 (f) and 2 (g) demonstrate the changes in the powder X-ray diffraction patterns during the heating of the mixture of $YH_3+H_2$ pressurized at ~200 and ~237 GPa for samples 2 and 1, respectively. Pulsed laser heating initiates the phase transformation of $YH_4$ into the $YH_6$ phase, and a subsequent transformation of $YH_6$ into the $YH_9$ phase at higher temperatures.

In order to estimate the stoichiometry of the newly synthesized yttrium hydrides, we studied $YH_3$ in a wide range of pressures. The experimentally obtained compressibility for the crystal structure of *fcc*-$YH_3$/$YD_3$ phase perfectly coincides with the theoretically calculated data[14,20] (Fig.3 (a)). Taking into account the volume occupied by a Y atom in its pure metallic phase at the same pressure[24], the additional volume caused by hydration, i.e. the difference between the volumes of $YH_3$ and Y, is 6.7 $Å^3$ at 150 GPa and 5.1 $Å^3$ at 215 GPa. Thus, the volume occupied by one hydrogen atom ($V_H$) depends on pressure and varies from 2.2 $Å^3$ to 1.7 $Å^3$ for pressures ranging between 150 and 215 GPa. These estimated values for $V_H$ are comparable to the one for the La-H system (1.9 $Å^3$ at 150−178 GPa)[2] and other metal-H systems[25-27]. Using the calculated values of 1.6 $Å^3$ for $V_H$ and 11.2 $Å^3$ for the volume of yttrium derived from the extrapolated data from the equation of state of metallic yttrium[24], the stoichiometry calculated from the experimental diffraction data for the new yttrium hydrides are $YH_{4.1}$, $YH_{5.8}$, and $YH_{8.8}$ at 237 GPa, respectively. The crystallographic structures of $YH_4$, $YH_6$, and $YH_9$, agree perfectly with the theoretical predictions[5,6]. The X-ray diffraction data for the volume of the crystal structure normalized with respect to the volume one Y atom as well as fragments of the crystal structures and coordination polyhedra for all experimentally found yttrium hydride phases are shown in Figure 3. The positions of the yttrium atoms were found directly from the diffraction data, whereas the hydrogen atoms were placed in the geometrically calculated positions based on the theoretical data[5].

In spite of the very good agreement between the predictions and the experimental crystallographic structures, the measured $T_c$s for the *Im-3m* $YH_6$ ($T_c$ ~227 K) and the *P6_3/mmc* $YH_9$ ($T_c$ ~243 K) phases are markedly lower than the predicted values of 251−264 K for $YH_6$[6] and 253−276 K for $YH_9$[5]. Thus, only the *fcc*-$YH_{10}$ phase can be expected to display RTSC with a predicted $T_c$ ~ 305−326 K[13]. However, we did not find $YH_{10}$, in spite of extensive trials with high-pressure synthesis up to 237 GPa. Still there is a possibility that this phase can be synthesized at higher pressures and temperatures. According to some predictions, the *fcc*-YH$_{10}$ phase is dynamically stable starting from 226 GPa[15] or 220 GPa[13]. However, other calculations[5] suggest that the *fcc*-YH$_{10}$ cannot be stabilized even at pressures as high as 400 GPa. Instead, the hexagonal YH$_9$ is energetically more favorable and lies on both convex hulls of formation enthalpy and internal energy, while YH$_{10}$ has a higher formation enthalpy and lies above the convex hull by 24 meV/atom at 400 GPa that is associated to 1100 K[5]. The synthesis attempt of *fcc*-YH$_{10}$ under higher pressures and temperatures is in progress.

**Methods**

To synthesize the initial YH$_3$ and YD$_3$, yttrium metal of 99.9% purity was first annealed in a vacuum of about 10$^{-3}$ Torr at 400 °C for 4 h, and then exposed to hydrogen (or deuterium) gas at a pressure of about 100 bars at 200 °C for 24 h in a high-pressure Sievert-type apparatus. The sample treatment was done in an Ar glovebox to prevent oxidation. The reaction products were YH$_{2.92(5)}$ and YD$_{2.87(5)}$ as indicated by their weight. We will further refer to these samples as YH$_3$ and YD$_3$ for brevity. The samples were analyzed through XRD with an Empyrean diffractometer at ambient conditions under a Kapton film. The lattice parameters of YH$_3$ and YD$_3$ were in reasonable agreement with the available data[28].

In the diamond anvil cells (DACs), we typically synthesized yttrium hydride via a direct reaction between yttrium (Alfa Aesar 99.9%) or YH$_3$ and hydrogen (99.999%) at high pressures. For that, a piece of Y or YH$_3$ was placed into a hole drilled in an insulating gasket. The process of synthesis is the same as the one followed for lanthanum hydride[2]. The pressure, pulsed laser heating temperature, and the amount of hydrogen surrounding the sample determined the composition of the yttrium hydrides. Superhydrides were synthesized only under an evident excess of hydrogen and high enough pressure. For the thermal treatment, one-sided pulsed radiation from a YAG laser was focused onto a spot having a diameter of 10 µm. Some samples were prepared not from elemental yttrium as the starting material but from YH$_3$ which was synthesized and analyzed as described above. One of the advantages of this method is to initially have a higher hydrogen content. To determine the isotope effect, we substituted hydrogen by deuterium (99.75% purity).

Typically, the diamonds used in the DACs had a culet with a diameter of 20−35 µm and were beveled at 8° to a diameter of about 250 µm. A toroidal profile was machined at each culet by a focused beam of xenon ions (FERA3 TESCAN). Tantalum electrodes were sputtered onto the surface of one of the diamond anvils in a van der Pauw four-probe geometry and were covered with gold. A metallic gasket

(T301 stainless steel) was electrically separated from the electrodes by an insulating layer (a mixture of epoxy and $CaF_2$, MgO, $CaSO_4$, cBN or $Al_2O_3$). The typical sample size was 5−10 µm.

We present resistance measurements upon warming the DACs as it yields a more accurate temperature reading: the cell was warmed up slowly (0.2 K min$^{-1}$) under an isothermal environment (no coolant flow). The temperature was measured with an accuracy of about 0.1 K by a Si diode thermometer attached to the DAC. All electrical transport measurements were performed with the electrical current set in the range of $10^{-5}$-$10^{-3}$ A. The pressure was measured through the $H_2$ ($D_2$) vibron scale[29] if such a vibron can be observed, or otherwise from the diamond Raman edge scale[30]. The $T_c$ was determined from the onset of superconductivity – at the point of apparent deviation from the temperature dependence of the resistance in the normal state metallic behavior.

We used three types of DACs. In particular, DACs with diameters of 20 mm and 8.8 mm were made of nonmagnetic materials, suitable for measurements under magnetic fields using a 9 T Quantum Design Physical Property Measurement System (PPMS). The X-ray diffraction measurements were done with wavelengths of 0.3344 Å and 0.2952 Å, an X-ray spot size ~3×3 µm, and Pilatus 1M CdTe detector at the beamline 13-IDD at GSECARS, Advanced Photon Source, Argonne National Laboratory (Chicago). Primary processing and integration of the powder patterns were made using the Dioptas software[31]. The Indexing and refinement were done with GSAS and EXPGUI packages[32].

**Figure captions**

**Figure 1. Superconducting transitions in yttrium hydrides.**
(a) The temperature dependence of resistance for the *Im-3m* $YH_6$ phase (blue curve, sample 1) and the $P6_3/mmc$ $YH_9$ phase (red curve, sample 2). Samples 1 and 2 were synthesized from $YH_3$ under hydrogen atmosphere. The configuration of the measurements is shown in the inset. After 3 weeks of maintaining sample 1 at 237 GPa and at room temperature, a sharp transition appeared, indicating a superconducting transition with an onset of $T_c$= 227 K – blue curve. This transition corresponds to the *Im-3m* $YH_6$ phase (see the details of the structure in Fig.2 (d) and 2 (f)). After keeping sample 2 at 201 GPa and at room temperature for ~ 1 month, with the aid of pulsed laser heating, the $P6_3/mmc$ $YH_9$ phase was observed with a $T_c$ =243 K (see Fig.2 (b) and 2 (g) for the structural characterization).
(b) The pressure dependence of $T_c$ for the different samples belonging to the $P6_3/mmc$ $YH_9$ phase: S1−filled magenta circles−sample 1 (after laser heating), S2 –filled black circles – sample 2 (after laser heating), S3−filled red circles–sample 3. Details concerning the synthesis of the samples can be found in the SM Table 1. Data shown by open black circles were obtained by decreasing the pressure of sample 2.

The open magenta circles correspond to the decompression of sample 1. The dotted line is a guide to the eyes.

(c) The pressure dependence of $T_c$ for the different samples belonging to the *Im-3m* YH$_6$ phase: S1−filled magenta squares−sample 1 (before laser heating), S2−filled black squares with main the superconducting transition at 210 K, smaller open black squares represent a small drop in resistance at 220 K−sample 2 (before laser heating), S4−filled olive squares–sample 4. The details concerning sample synthesis can be found in SM Table 1. Samples 1 and 2 crystallize in the *Im-3m* YH$_6$ and in the *I4/mmm* YH$_4$ phases, with the $T_c$ dominated by the *Im-3m* YH$_6$ phase as indicated by the black polygon. Sample 4 crystallized in the *Im-3m* YH$_6$ and *I4/mmm* YH$_4$ phases in addition to a complex unidentified phase (or a mixture of phases).

(d) Superconducting transition for sample 5 under an external magnetic field. This panel displays the electrical resistance as a function of the temperature under applied magnetic fields up to 9 T. For sample 5, $R(T)$ was measured through a three terminal method, and the resulting background was subsequently subtracted. The temperature dependence of the observed upper critical fields (inset) was obtained from the data shown in Fig. 1 (d). An extrapolation to the lowest temperatures yields an ~110 T for the upper critical magnetic field in the limit of zero temperatures.

**Figure 2. X-ray powder diffraction analysis of the synthesized yttrium hydrides through Rietveld refinement of the crystal structures.** Black crosses are the experimental points, solid red lines are the Rietveld fits, and the blue solid lines represent the residuals. The blue, magenta and green ticks indicate the diffraction peaks corresponding to the tetragonal YH$_4$, cubic YH$_6$ and hexagonal YH$_9$ phases, respectively. The used X-ray wavelengths (λ, Å) are indicated in the top right corners of each plot. The relative contribution between phases in a Rietveld refinement comprising two different crystalline phases is summarized at the bottom right corner of each plot. Fit parameters for each refinement are shown under the blue residual graph.

(a) Rietveld refinement for the *I4/mmm* YH$_4$ crystal structure found as a pure single-phase in some regions of sample 4 subjected to 183 GPa.

(b) Rietveld refinement for the *Im-3m* YH$_6$ crystal structure found in sample 2 under 201 GPa which was kept at room temperature for 1 month.

(c) Rietveld refinement for the *P6$_3$/mmc* YH$_9$ crystallographic structure found in sample 1 under 237 GPa after annealing for five cycles of very subtle pulsed laser heating (the temperature of each cycle was kept below 700 K).

(d) Cake representation of the raw X-ray diffraction patterns collected for sample 1, which was kept at room temperature for 3 weeks. The three broad and highly spotted lines correspond to the *Im-3m* phase of YH$_6$ (this raw pattern corresponds to the black powder pattern in (f)).

(e) The same cake representation of sample 1 after five cycles of very subtle heating up to temperatures below 700 K. The new narrow and dashed lines appearing after the heat treatment and correspond to the $P6_3/mmc$ phase of $YH_9$ (this raw pattern corresponds to the green powder pattern in (f)).

(f) The changes in the powder patterns of sample 1 after successive cycles of very subtle heating (the temperature of each cycle is maintained below 700 K). The black powder pattern before any pulsed laser heating corresponds to the nearly pure $Im\text{-}3m$ phase of $YH_6$. Each heating cycle results in the conversion of $YH_6$ into the higher $YH_9$ hydride, as indicated by the appearance of new diffraction peaks related to the $P6_3/mmc$ lattice. Reflections from the $Im\text{-}3m$ $YH_6$ and $P6_3/mmc$ $YH_9$ phases are marked by $c$ and $h$, respectively.

(g) Changes in the powder diffraction patterns after several cycles of pulsed laser heating for sample 2. The laser power and the corresponding temperature at each successive heating cycle was continuously increased up to 1750(15) K. At the beginning, the sample mainly consists of the $Im\text{-}3m$ $YH_6$ and $I4/mmm$ $YH_4$ phases. Each heating cycle that maintains the temperature below ~1300 K initiates the transformation of the $YH_4$ phase into $YH_6$. Successive heating cycles to higher temperatures initiate the transformation of $YH_6$ into the $YH_9$ phase.

**Figure 3. Equation of state and crystal structures for the different $YH_x$ ($0 \leqslant x \leqslant 9$) phases.**

(a) Unit cell volume normalized to the volume of one Y atom as a function of the pressure for the different yttrium hydrides as well as for pure yttrium. Data related to the same given phase are indicated by the red polygons. Hexagons, stars, squares, triangles, and filled circles correspond to $P6_3/mmc$ $YH_9$, $Im\text{-}3m$ $YH_6$, $I4/mmm$ $YH_4$, $Fm\text{-}3m$ $YH_3$, and distorted $Fm\text{-}3m$ ($hR24$ and $C2/m$) yttrium, respectively. Red, blue, green, magenta, purple, grey, navy, orange, azure, dark green, light grey, dark yellow, brown, dark violet, black, light pink, light brown markers correspond to samples 1, 2, 3, 4, 5, 7, 8, 9, 10, 11, 12, 13, 14, 15, 16, 17, and 18, respectively (see the SM Table 1). Filled black circles correspond to the experimental data for metallic yttrium[24]. Open black symbols correspond to the theoretically predicted crystallographic structures: open stars for $YH_6$[5,6], open squares for $YH_4$[6], and open triangles for $YH_3$[14,20].

(b) Crystallographic structure for the different yttrium hydrides found in the experiment based on the X-ray diffraction data from the lattice of the heavier yttrium atoms. Hydrogen atoms were geometrically placed using their positions according to the theoretically predicted structures[5]. Yttrium and hydrogen atoms are coloured in blue and grey, respectively. The unit cell is indicated by red lines. The coordination polyhedra (cages) and building polygons for these structures ($YH_3$, $YH_4$, $YH_6$, $YH_9$) are indicated by the light gray lines with the corresponding compositions shown in the row immediately below each structure.

**Acknowledgements.**

M.E. is thankful to the Max Planck community for the invaluable support, and U. Pöschl for the constant encouragement. L.B. is supported by DOE−BES through award DE-SC0002613. S.M. acknowledges support from the FSU Provost Postdoctoral Fellowship Program. The NHMFL acknowledges support from the U.S. NSF Cooperative Grant No. DMR−1644779, and the State of Florida. Portions of this work were performed at GeoSoilEnviro CARS (The University of Chicago, Sector 13), Advanced Photon Source (APS), Argonne National Laboratory. GeoSoilEnviro CARS is supported by the National Science Foundation−Earth Sciences (EAR−1634415) and Department of Energy−GeoSciences (DE−FG02−94ER14466). This research used resources of the Advanced Photon Source, a U.S. Department of Energy (DOE) Office of Science User Facility operated for the DOE Office of Science by Argonne National Laboratory under Contract No. DE−AC02−06CH11357.

**Correspondence and requests for materials** should be addressed to M.E.

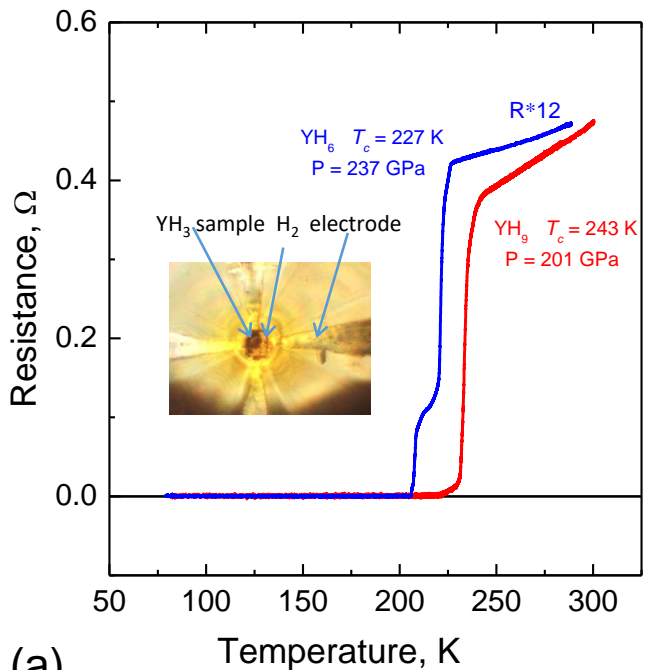
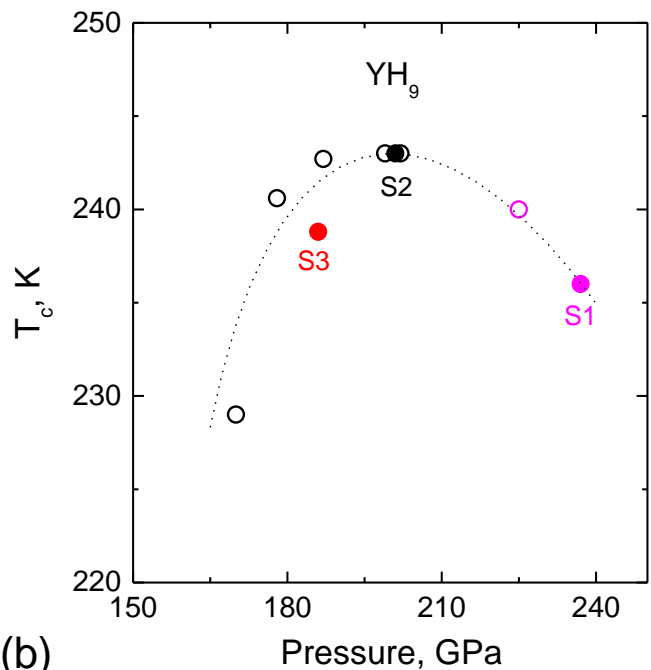
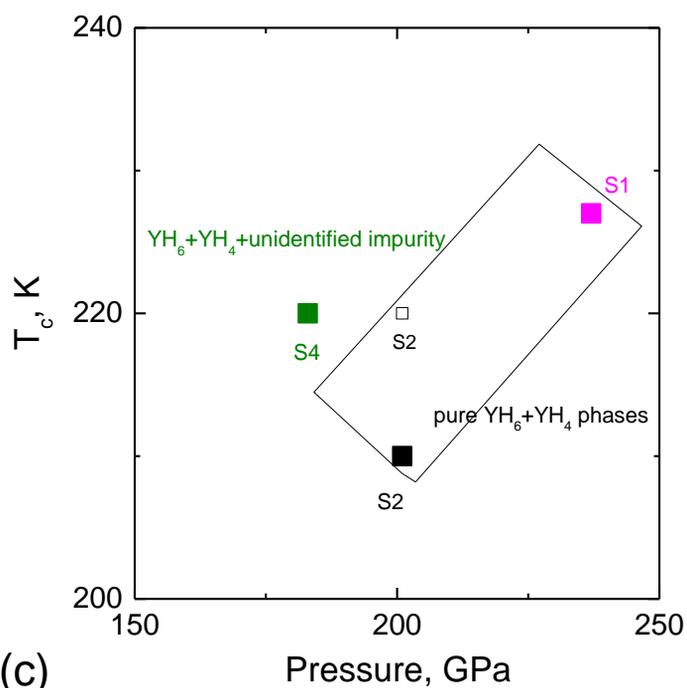
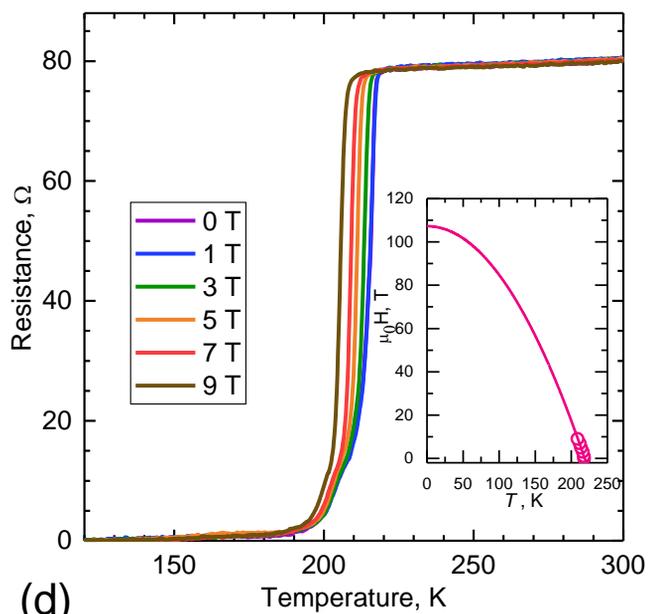

**Figure 1. Superconducting transitions in yttrium hydrides.**
(a) The temperature dependence of resistance for the *Im-3m* $YH_6$ phase (blue curve, sample 1) and the $P6_3/mmc$ $YH_9$ phase (red curve, sample 2). Samples 1 and 2 were synthesized from $YH_3$ under hydrogen atmosphere. The configuration of the measurements is shown in the inset. After 3 weeks of maintaining sample 1 at 237 GPa and at room temperature, a sharp transition appeared, indicating a superconducting transition with an onset of $T_c$= 227 K – blue curve. This transition corresponds to the *Im-3m* $YH_6$ phase (see the details of the structure in Fig.2 (d) and 2 (f)). After keeping sample 2 at 201 GPa and at room temperature for ~ 1 month, with the aid of pulsed laser heating, the $P6_3/mmc$ $YH_9$ phase was observed with a $T_c$=243 K (see Fig.2 (b) and 2 (g) for the structural characterization).
(b) The pressure dependence of $T_c$ for the different samples belonging to the $P6_3/mmc$ $YH_9$ phase: S1–filled magenta circles–sample 1 (after laser heating), S2 –filled black circles – sample 2 (after laser heating), S3–filled red circles–sample 3. Details concerning the synthesis of the samples can be found in the SM Table 1. Data shown by open black circles were obtained by decreasing the pressure of sample 2. The open magenta circles correspond to the decompression of sample 1. The dotted line is a guide to the eyes.
(c) The pressure dependence of $T_c$ for the different samples belonging to the *Im-3m* $YH_6$ phase: S1–filled magenta squares–sample 1 (before laser heating), S2–filled black squares with main the superconducting transition at 210 K, smaller open black squares represent a small drop in resistance at 220 K–sample 2 (before laser heating), S4–filled olive squares–sample 4. The details concerning sample synthesis can be found in SM Table 1. Samples 1 and 2 crystallize in the *Im-3m* $YH_6$ and in the *I4/mmm* $YH_4$ phases, with the $T_c$ dominated by the *Im-3m* $YH_6$ phase as indicated by the black polygon. Sample 4 crystallized in the *Im-3m* $YH_6$ and *I4/mmm* $YH_4$ phases in addition to a complex unidentified phase (or a mixture of phases).
(d) Superconducting transition for sample 5 under an external magnetic field. This panel displays the electrical resistance as a function of the temperature under applied magnetic fields up to 9 T. For sample 5, *R* (*T*) was measured through a three terminal method, and the resulting background was subsequently subtracted. The temperature dependence of the observed upper critical fields (inset) was obtained from the data shown in Fig. 1 (d). An extrapolation to the lowest temperatures yields an ~110 T for the upper critical magnetic field in the limit of zero temperatures.

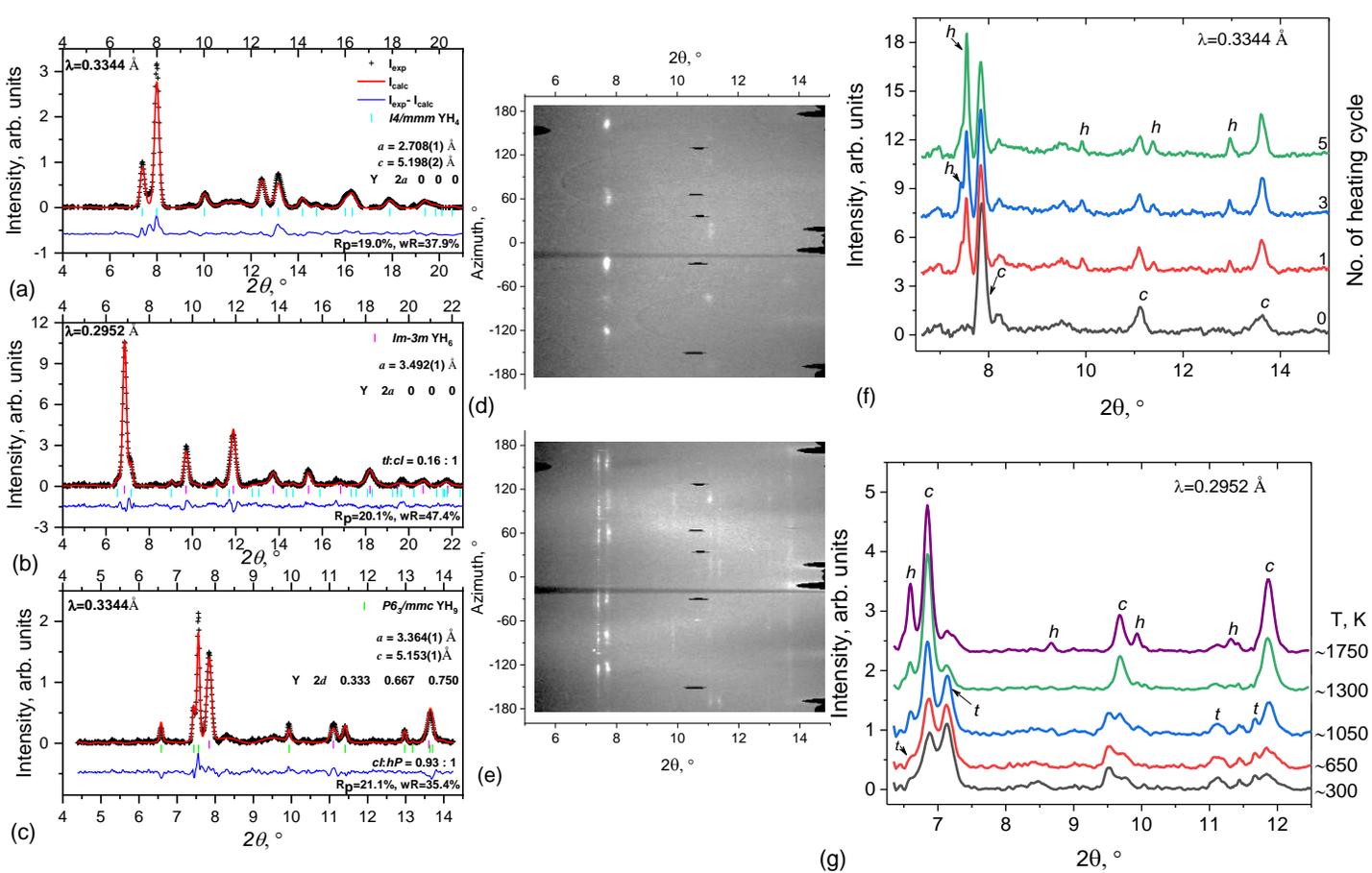

**Figure 2. X-ray powder diffraction analysis of the synthesized yttrium hydrides through Rietveld refinement of the crystal structures.** Black crosses are the experimental points, solid red lines are the Rietveld fits, and the blue solid lines represent the residuals. The blue, magenta and green ticks indicate the diffraction peaks corresponding to the tetragonal $YH_4$, cubic $YH_6$ and hexagonal $YH_9$ phases, respectively. The used X-ray wavelengths (λ, Å) are indicated in the top right corners of each plot. The relative contribution between phases in a Rietveld refinement comprising two different crystalline phases is summarized at the bottom right corner of each plot. Fit parameters for each refinement are shown under the blue residual graph.
(a) Rietveld refinement for the $I4/mmm$ $YH_4$ crystal structure found as a pure single-phase in some regions of sample 4 subjected to 183 GPa.
(b) Rietveld refinement for the $Im-3m$ $YH_6$ crystal structure found in sample 2 under 201 GPa which was kept at room temperature for 1 month.
(c) Rietveld refinement for the $P6_3/mmc$ $YH_9$ crystallographic structure found in sample 1 under 237 GPa after annealing for five cycles of very subtle pulsed laser heating (the temperature of each cycle was kept below 700 K).
(d) Cake representation of the raw X-ray diffraction patterns collected for sample 1, which was kept at room temperature for 3 weeks. The three broad and highly spotted lines correspond to the $Im-3m$ phase of $YH_6$ (this raw pattern corresponds to the black powder pattern in (f)).
(e) The same cake representation of sample 1 after five cycles of very subtle heating up to temperatures below 700 K. The new narrow and dashed lines appearing after the heat treatment and correspond to the $P6_3/mmc$ phase of $YH_9$ (this raw pattern corresponds to the green powder pattern in (f)).
(f) The changes in the powder patterns of sample 1 after successive cycles of very subtle heating (the temperature of each cycle is maintained below 700 K). The black powder pattern before any pulsed laser heating corresponds to the nearly pure $Im-3m$ phase of $YH_6$. Each heating cycle results in the conversion of $YH_6$ into the higher $YH_9$ hydride, as indicated by the appearance of new diffraction peaks related to the $P6_3/mmc$ lattice. Reflections from the $Im-3m$ $YH_6$ and $P6_3/mmc$ $YH_9$ phases are marked by $c$ and $h$, respectively.
(g) Changes in the powder diffraction patterns after several cycles of pulsed laser heating for sample 2. The laser power and the corresponding temperature at each successive heating cycle was continuously increased up to 1750(15) K. At the beginning, the sample mainly consists of the $Im-3m$ $YH_6$ and $I4/mmm$ $YH_4$ phases. Each heating cycle that maintains the temperature below ~1300 K initiates the transformation of the $YH_4$ phase into $YH_6$. Successive heating cycles to higher temperatures initiate the transformation of $YH_6$ into the $YH_9$ phase.

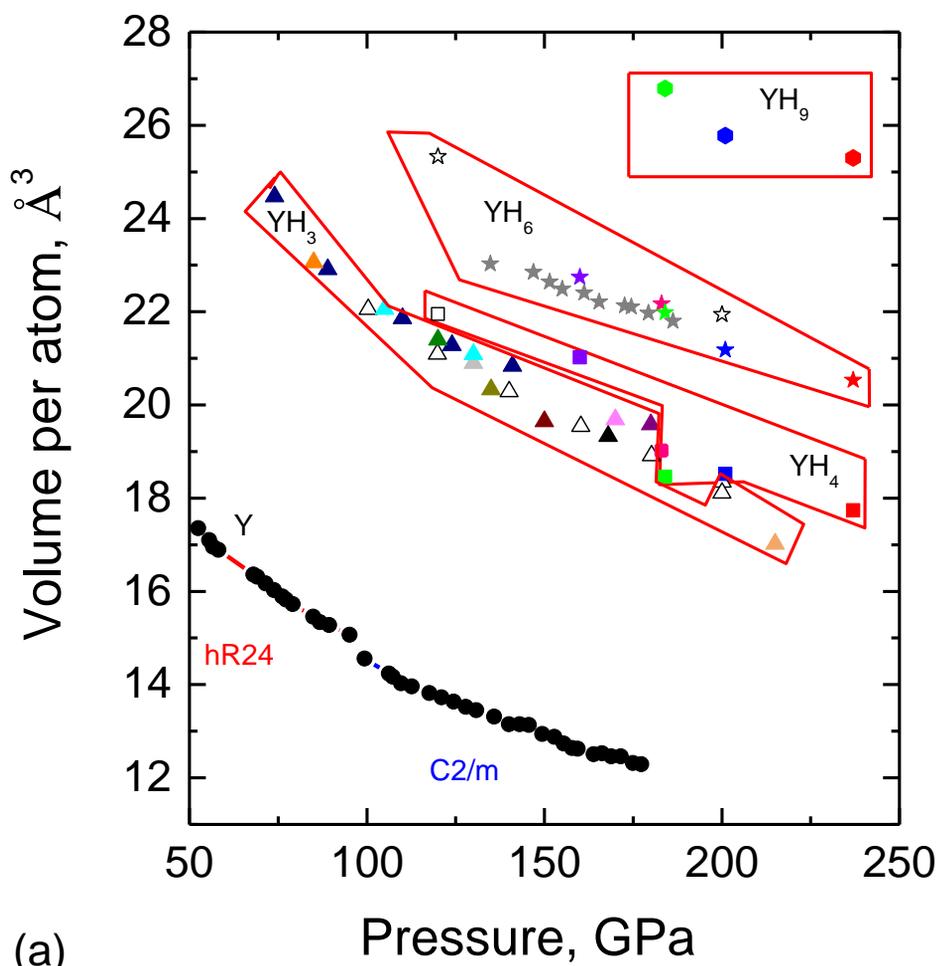

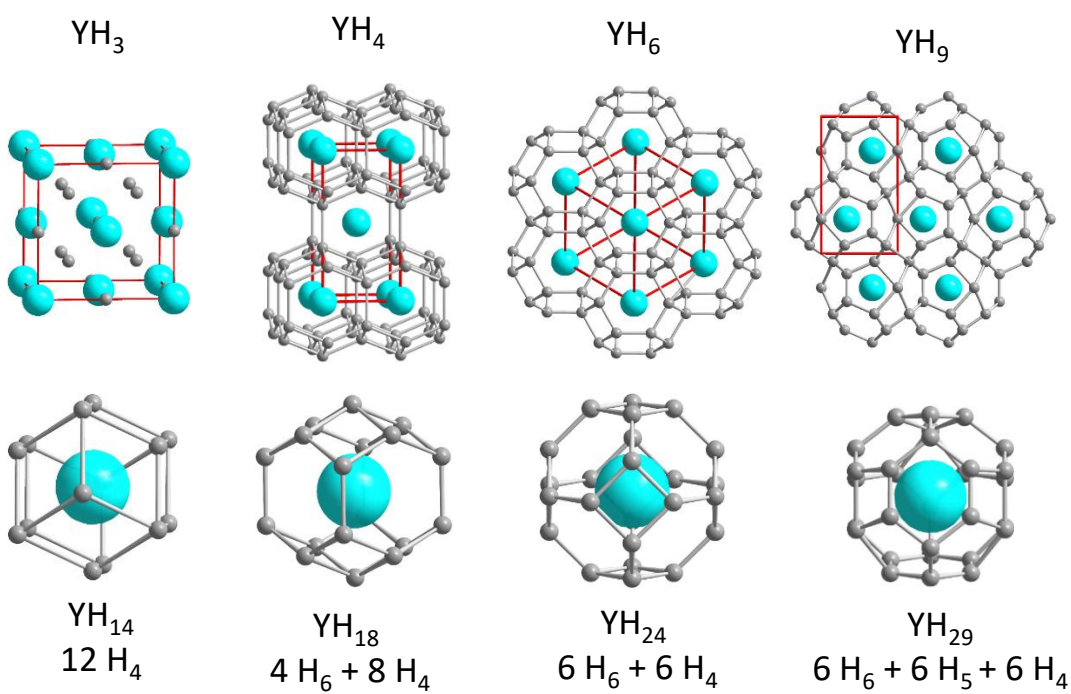

**Figure 3. Equation of states and crystal structures of different YH$_x$ (0 ≤ x ≤ 9) phases.**
(a) Unit cell volume normalized to the volume of one Y atom as a function of the pressure for the different yttrium hydrides as well as for pure yttrium. Data related to the same given phase are indicated by the red polygons. Hexagons, stars, squares, triangles, and filled circles correspond to *P6$_3$/mmc* YH$_9$, *Im-3m* YH$_6$, *I4/mmm* YH$_4$, *Fm-3m* YH$_3$, and distorted *Fm-3m* (*hR24* and *C2/m*) yttrium, respectively. Red, blue, green, magenta, purple, grey, navy, orange, azure, dark green, light grey, dark yellow, brown, dark violet, black, light pink, light brown markers correspond to samples 1, 2, 3, 4, 5, 7, 8, 9, 10, 11, 12, 13, 14, 15, 16, 17, and 18, respectively (see the SM Table 1). Filled black circles correspond to the experimental data for metallic yttrium[24]. Open black symbols correspond to the theoretically predicted crystallographic structures: open stars for YH$_6$[5,6], open squares for YH$_4$[6], and open triangles for YH$_3$[14,20].
(b) Crystallographic structure for the different yttrium hydrides found in the experiment based on the X-ray diffraction data from the lattice of the heavier yttrium atoms. Hydrogen atoms were geometrically placed using their positions according to the theoretically predicted structures[5]. Yttrium and hydrogen atoms are coloured in blue and grey, respectively. The unit cell is indicated by red lines. The coordination polyhedra (cages) and building polygons for these structures (YH$_3$, YH$_4$, YH$_6$, YH$_9$) are indicated by the light gray lines with the corresponding compositions shown in the row immediately below each structure.

| Sample | Synthesis conditions | Electrical measurements | X-ray diffraction |
|---|---|---|---|
| 1 (1010) | $YH_3 + H_2$. Pressurized to 255 (or 237 according to $H_2$ vibron) GPa and kept for 3 weeks at room temperature. | $T_c$ = 227 K Fig.1c, magenta square | **Im-3m** $YH_6$, dominant phase: $a$=3.457(1) Å, $V$=41.3(1) Å$^3$; **I4/mmm** $YH_4$: $a$=2.616 Å, $c$=5.184 Å, $V$=35.5 Å$^3$. |
| | The same sample after 5 subtle laser heating cycles (the temperature of each cycle is below 700 K, no visible glowing). | $T_c$ = 237 K Fig.1b, filled magenta circle | **P6$_3$/mmc** $YH_9$ phase: $a$=3.364(1) Å, $c$=5.153(1) Å, $V$=50.5(1) Å$^3$; with remains of **Im-3m** $YH_6$ and **I4/mmm** $YH_4$. |
| 2 (Play) | $YH_3 + H_2$. Pressurized to 238 (or 201 according to $H_2$ vibron) GPa and kept for 1 months at room temperature. | Fig.1a, blue curve, Fig.1c, black square $T_c$ = 210 K | **Im-3m** $YH_6$, dominant phase: $a$=3.492(1) Å, $V$=42.6(1) Å$^3$; with **I4/mmm** $YH_4$: $a$=2.656(1) Å, $c$=5.190(8) Å, $V$=36.6(1) Å$^3$. |
| | The same sample after tens of laser heating cycles up to 2000(10) K. | Fig.1a, red curve, Fig.1b, filled black circle $T_c$ = 243 K | **P6$_3$/mmc** $YH_9$ phase: $a$=3.406 Å, $c$=5.210 Å, $V$=52.3 Å$^3$; with remains of **Im-3m** $YH_6$ and traces of **I4/mmm** $YH_4$. |
| 3 (H120) | $Y + H_2$. Pressurized to 235 (or 184 according to $H_2$ vibron) GPa. Heated by pulsed laser. | Fig.1b, filled red circle $T_c$ = 239 K | **I4/mmm** $YH_4$, dominant phase: $a$=2.666 Å, $c$=5.194 Å, $V$=36.9 Å$^3$; **Im-3m** $YH_6$ phase: $a$=3.529 Å, $V$=43.9 Å$^3$; **P6$_3$/mmc** $YH_9$ phase: $a$=3.432 Å, $c$=5.251 Å, $V$=53.6 Å$^3$. |
| 4 (YH1) | $Y + H_2$. Pressurized to 189 (or 183 according to $H_2$ vibron) GPa. Heated by pulsed laser. | Fig.1c, olive square $T_c$ = 220 K | The mixture of **I4/mmm** $YH_4$: $a$=2.708(1) Å, $c$=5.197(3) Å, $V$=38.1(1) Å$^3$; and **Im-3m** $YH_6$ phase: $a$=3.542(2) Å, $V$=44.4(1) Å$^3$. |
| 5 (Q112) | $Y + H_2$. Pressurized to 185 (or 160 according to $H_2$ vibron) GPa. Heated by pulsed laser. | Fig.1 d $T_c$ ~ 214 K | The mixture of **I4/mmm** $YH_4$: $a$=2.766 Å, $c$=5.494 Å, $V$=42.0 Å$^3$; and **Im-3m** $YH_6$ phase: $a$=3.570 Å, $V$=45.5 Å$^3$; and unidentified impurity(s). |
| 6 (Q142) | $YD_3 + D_2$. Pressurized to 194 (or 173 according to $D_2$ vibron) GPa. Heated by pulsed laser. | $T_c$ ~ 170 K | |
| 7 (Q121) | $YD_3 + D_2$. Pressurized to 198 (or 170 according to $D_2$ vibron) GPa. Heated by pulsed laser. | w/o electrodes | The mixture of **Im-3m** $YD_6$ phase: $a$=3.536(1) Å, $V$=44.2(1) Å$^3$; and unidentified impurity(s). |
| | The same sample on decompression down to 135 GPa (then the cell was broken). | | The mixture of **Im-3m** $YD_6$ phase: $a$=3.585(6) Å, $V$=46.1(2) Å$^3$ (at 135 GPa); and unidentified impurity(s). |
| 8 (cell28) | $Y + H_2$. Pressurized to 23 GPa. Structural changes were followed by X-ray diffraction on compression up to 131 GPa. | w/o electrodes | **Fm-3m** $YH_3$ phase: $a$=4.928 Å, $V$=119.7 Å$^3$ (at 23 GPa). |
| | The same sample after laser heating at 131 GPa and compression up to 140 GPa. | | Unidentified phase(s); and remains of **Fm-3m** $YH_3$ phase: $a$=4.41 Å, $V$=85.8 Å$^3$ (at 140 GPa). |
| 9 (CT1) | $Y + H_2$. Pressurized to 85 GPa and heated by pulsed laser. | w/o electrodes | **Fm-3m** $YH_3$ phase: $a$=4.536 Å, $V$=93.3 Å$^3$. |
| 10 (QL11) | $Y + H_2$. Pressurized to 105 GPa and slightly heated by pulsed laser. | w/o electrodes | **Fm-3m** $YH_3$ phase: $a$=4.452(1) Å, $V$=88.2(1) Å$^3$. |
| | The same sample heated by pulsed laser several times at 105 and 130 GPa up to ~1800 K. | | Unidentified phase(s); and remains of **Fm-3m** YH$_3$ phase: $a$=4.388(1) Å, $V$=84.5(1) Å$^3$ at 130 GPa. |
| 11 (Q3) | YH$_3$ + H$_2$. Pressurized to 120 GPa and slightly heated by pulsed laser. | w/o electrodes | Mainly **Fm-3m** YH$_3$ phase: $a$=4.407(1) Å, $V$=85.6(1) Å$^3$; and traces of unidentified impurity(s). |
| 12 (W5) | YH$_3$. Pressurized to 130 GPa and slightly heated by pulsed laser. | Metallic behavior on cooling with a residual resistance of ~0.04 Ohm at 5 K. | The mixture of **Fm-3m** YH$_3$: $a$=4.373(6) Å, $V$=83.6(4) Å$^3$; and **Fm-3m** YH phase: $a$=3.986(2) Å, $V$=63.4(1) Å$^3$. |
| 13 (W7) | YD$_3$. Pressurized to 135 GPa and slightly heated by pulsed laser. | Metallic behavior on cooling with a residual resistance of ~0.02 Ohm at 5 K. | The mixture of **Fm-3m** YD$_3$: $a$=4.332(1) Å, $V$=81.3(1) Å$^3$; and **Fm-3m** YD phase: $a$=3.975(1) Å, $V$=62.8(1) Å$^3$. |
| 14 (Q1) | YH$_3$ + H$_2$. Pressurized to 150 GPa and heated by pulsed laser up to ~1800 K. | w/o electrodes | Unidentified phase(s); and remains of **Fm-3m** YH$_3$ phase: $a$=4.284(3) Å, $V$=78.6(2) Å$^3$. |
| 15 (G2) | YD$_3$. Pressurized to 170 GPa and slightly heated by pulsed laser. | Metallic behavior on cooling with a residual resistance of ~0.5 Ohm at 5 K. | The mixture of **Fm-3m** YD$_3$: $a$=4.279(2) Å, $V$=78.3(2) Å$^3$; and **Fm-3m** YD phase: $a$=3.876(1) Å, $V$=58.2(1) Å$^3$. |
| 16 (W10) | YD$_3$. Pressurized from 4 to 168 GPa at ~100 K and then warmed. | w/o electrodes | **Fm-3m** YD$_3$ phase: $a$=4.260 Å, $V$=77.3 Å$^3$; and traces of **Fm-3m** YD phase. |
| 17 (G1) | YH$_3$. Pressurized to 170 GPa and heated by pulsed laser. | Metallic behavior on cooling with a residual resistance of ~0.05 Ohm at 5 K. | The mixture of **Fm-3m** YH$_3$: $a$=4.286(2) Å, $V$=78.8(1) Å$^3$; and **Fm-3m** YH phase: $a$=3.901(1) Å, $V$=59.4(1) Å$^3$. |
| 18 (CM01) | Y + H$_2$. Pressurized to 215 (or 180 according to H$_2$ vibron) GPa. Heated by pulsed laser up to ~1800 K. | w/o electrodes | Unidentified phase(s); and remains of **Fm-3m** YH$_3$ phase: $a$=4.083 Å, $V$=68.1 Å$^3$. |